# Locally Weighted Regression with different Kernel Smoothers for Software Effort Estimation

Yousef Alqasrawi[1], Mohammad Azzeh[2,] Yousef Elsheikh[3]

[1,3]Faculty of Information Technology, Applied Science Private University, Amman, Jordan

[2]Department of Data Science, Princess Sumaya University for Technology, Amman, Jordan

[1]y_alqasrawi@asu.edu.jo, [2]m.azzeh@psut.edu.jo, [3]y_elsheikh@asu.edu.jo

Abstract. Estimating software effort has been a largely unsolved problem for decades. One of the main reasons that hinders building accurate estimation models is the often heterogeneous nature of software data with a complex structure. Typically, building effort estimation models from local data tends to be more accurate than using the entire data. Previous studies have focused on the use of clustering techniques and decision trees to generate local and coherent data that can help in building local prediction models. However, these approaches may fall short in some aspect due to limitations in finding optimal clusters and processing noisy data. In this paper we used a more sophisticated locality approach that can mitigate these shortcomings that is Locally Weighted Regression (LWR). This method provides an efficient solution to learn from local data by building an estimation model that combines multiple local regression models in *k*-nearest-neighbor based model. The main factor affecting the accuracy of this method is the choice of the kernel function used to derive the weights for local regression models. This paper investigates the effects of choosing different kernels on the performance of Locally Weighted Regression of a software effort estimation problem. After comprehensive experiments with 7 datasets, 10 kernels, 3 polynomial degrees and 4 bandwidth values with a total of 840 Locally Weighted Regression variants, we found that: 1) Uniform kernel functions cannot outperform non-uniform kernel functions, and 2) kernel type, polynomial degrees and bandwidth parameters have no specific effect on the estimation accuracy. In other words, no change in bandwidth or degree values occurred with a significant difference in kernel rankings. In short, Locally Weighted Regression methods with Triweight or Triangle kernel can perform better than more complex kernels. Hence, we encourage non-uniform kernel methods as smoother function with wide bandwidth and small polynomial degree.

1. Introduction

Software effort estimation remains a largely unsolved problem due to the nature of software data that is often heterogenous with complex structure [1]–[5]. Most of the machine learning methods that are used for software effort estimation are global learning methods or global function approximations aimed at minimizing the global loss functions [1], [6], [7]. The solutions obtained by global methods cannot provide a sufficiently good approximation. In contrast, Local learning methods divide the global learning problem into multiple simpler local learning models which reasonably increase predictive performance [1], [3], [8]. This can be achieved by dividing the cost function into multiple independent local cost functions. In fact, the use of data locality to build prediction models has shown great interest within the research community as it helps to minimize model complexity, reduce bias and enhance accuracy [8][9][10][11]. Since effort estimation datasets tend to be rather small and heterogeneous, the locality approaches are likely to be more adequate and produce better accuracy than models that do not use locality [12][13]. Our focus in this paper is not on introducing a new locality method for effort estimation, but rather on examining current methods thoroughly and investigating their best configurations for obtaining local prediction models. Although many locality approaches have been successfully used with observable accuracy improvements (e.g. [1], [9], [12], [19], [20]), in this paper we want to study unexplored method in effort estimation that combines the benefit from both instance learning and data locality which is Locally Weighted Regression (LWR). Previous studies mainly focused on two locality approaches, namely, clustering techniques and decision trees to generate local and coherent data. Many studies have addressed the utility of clustering techniques for data locality such as [11], [14], [15]. Other studies have approached the locality in a different way, focusing on the use of in-company and cross-company dataset to build effort prediction models [10], [16]–[18]. Regardless of their good performance, these methods fall short due to limitations in identifying optimal clusters and processing noisy data. These limitations are addressed in LWR as 1) No need to search for optimal cluster numbers and 2) Very small weights are given to e ach irrelevant nearest instances. The LWR method builds multiple local regression models in *k*-nearest neighbors space [21]. The LWR is non-parametric learning where number of parameters grows with the number of training examples [21][22]. The basic idea of LWR is that instead of building a global model for the entire data, just only the neighboring data of the interest observation are taken to generate local model where the final prediction is made by the local function [22]. In other words, each near instance becomes a weighting factor based on the distance with target case such that the nearest instances are given higher weights than far away instances. To find the appropriate weights for the nearest neighbors, a kernel smoother function is usually used to smooth the final regression model [21], [22]. In literature, various kernel functions can be used with LWR which they can be classified into uniform and non-uniform kernel [23]. Uniform kernel treats all neighbors equally by producing equal weights for them [23]. Whereas non-uniform kernels treat each nearest neighbor differently by generating different weights for each instance. The size of kernel bandwidth and degree of regression model also have a major influence on the accuracy of final LWR model [22], [24].

Yet, although kernel functions can improve prediction performance of LWR, overall results seem to be quite mixed and inconsistent. We believe that there are three main reasons for this uncertainty: First, different studies employ different experimental procedures, choice of hyper-parameters, etc. which may render results not strictly comparable. Second, characteristic of the software effort datasets and its relationship with the predictive performance of LWR are unexplored. Third, the interaction between the choice of kernel methods and choice of bandwidth and polynomial degree is not well understood. Therefore, this paper examines the performance of different kernel smoother functions with multiple parameter settings (i.e., bandwidth and regression degree) for LWR effort estimation method. To help in understanding the impact of kernel functions there is a need to systematically explore the research questions below:

RQ1. Is there any evidence that non-uniform kernel functions can outperform uniform kernel functions for LWR?

- The purpose of this question is to examine whether the non-uniform kernel functions can significantly outperform uniform kernel function when used within LWR method.

RQ2. What are the interactions between: (*i*) data set dimensionality (*ii*) type of kernel (*iii*) polynomial degree and (*iv*) kernel bandwidth?
- The purpose of this question is to study relationship between different parameter settings of LWR method in order to give insight on the impact of each parameter on the overall performance of LWR method. The answer of this question can give us the relationship between dimension of datasets and favorite kernel functions. In addition, it can give us insight on the relationship between kernel functions and their associated parameters (degree and bandwidth).

To facilitate answering these questions, an extensive experimentation of 7 datasets, 4 evaluation criteria, 10 kernels, 3 polynomial degrees, 4 bandwidth values and a total of 840 LWR variants were performed. Since the number of experiments is dauntingly large, we use average aggregation of accuracies in addition to Scott-Knott analysis to present results properly. This will facilitate data exploration and analysis [4][25].

In summary, we found that using uniform kernel function with LWR cannot outperform non-uniform kernel functions, and kernel type, polynomial degrees and bandwidth parameters do not produce a definite effect on estimation accuracy. In other words, changing the bandwidth or degree values did not come with significant difference in the kernel rankings. Locally Weighted Regression methods with Triweight or Triangle kernel can perform better than much more complex kernels. Hence, we encourage using non-uniform kernel methods as smoother function with wide bandwidth and small polynomial degree.

The present paper is organized as follows: section 2 presents background about LWR and kernel smoother functions. Section 3 presents related work. Section 4 and 5 show the data and evaluation measures used. Section 6 presents research methodology. Section 7 presents results and discussion, and finally sections 8 and 9 ends with threats to validity of our work and show the conclusions.

## 2. Background

### 2.1 Locally Weighted Regression

Locally Weighted Regression is a type of nonparametric regression technique that is used to fit simple regression models to localized subsets of data [21]. This method builds multiple local regression models in $k$-nearest neighbors space as shown in Figure 1. LWR inherits much of the simplicity of linear least squares regression with the flexibility of nonlinear regression [22], [24]. In fact, one of the primary attractions of this method is that there is no need to fit data by global function, but rather using local functions based on subset of data. The LWR can be used for situations in which there is no suitable parametric regression form, and when there are outliers in the data for which a robust fitting method is necessary. However, LWR is also called lazy learning because the processing of the training data is shifted until a query point needs to be answered [26]. This approach makes LWR a very accurate function approximation method where it is easy to add new training points. The general form of the LWR is defined in equation 1. Two parameters are required for tuning LWR model which are weights of nearest neighbors and degree of polynomial function ($d$). Weights are usually generated by kernel function as shown in the next section, which we divide into uniform and non-uniform kernels. The degree parameter is usually determined by the expert who construct LWR method. The LWR is constructed in the way that at each point in the predictor variable, the nearest neighbor algorithm is consulted to sample the subset of data that is utilized for weighting instances [24], [26]. The data points that are near to the target point are given more weights than far data points. This process is accomplished by specifying the bandwidth or smoothing parameters which determines the amount of data used to fit each local regression model. Finally, it is recommended to use degree value from 1 to 3 to avoid overfitting [24].

$$f(x_0; \emptyset) = \sum_{i=1}^{k} (\emptyset_i \, x_i)^d \quad (1)$$

Where $k$ is number of nearest neighbors, $\emptyset$ is weight vector and $\emptyset_i$ is the weight assigned to each nearest instance which can be obtained from the kernel function $K(x_i, x_0)$, and $d$ is the degree of the polynomial, $x_i$ is the $i^{th}$ input variable. $x_0$ is the query example.

During training, the weights of the training examples are updated ($\emptyset^*$) using equation:

$$\emptyset^* = \sum_{i=1}^{n} L(y_i, f(x_0; \emptyset)) \, K(x_i, x_0) \quad (2)$$

Where $L(y_i, f(x_0; \emptyset)) = (y_i - x_i^T \emptyset)^2$ is squared loss function. $y_i$ is the actual output of the $i^{th}$ observation. $K(x_i, x_0)$ denotes to kernel function.

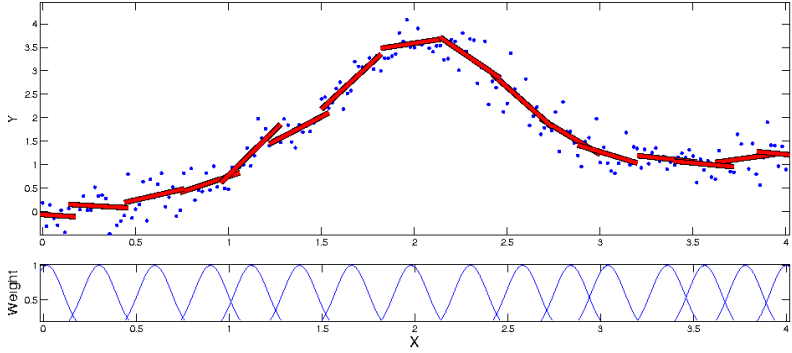

Figure 1. LWR model construction from local regression models. The red bars are local linear regression models that are constructed from local data using weights that are generated from kernel functions as shown in the plots beneath LWR model.

## 2.2 Choice of Kernel

As mentioned in the previous section, the LWR method requires kernel smoother function to help in producing the most appropriate weights for nearest neighbors of the target data points. We study kernel functions with LWR since, if the effort and project size data correspond to a particular distribution, then it would seem wise to bias that sampling by that distribution [23]. The general definition of kernel function is shown in equation 3 which is remarkably influenced by the bandwidth of nearest neighbors. The weights are generated by different types of kernels so that closer points are given higher weights. The estimated function is smooth, and the level of smoothness is set by the bandwidth. The bandwidth ($b$) is the proportion of data used in each fit which is usually set to $1/\sqrt{n}$ where $n$ is number of all instances. Thus, the subset of the data in each local model consists of $nb$ of nearest neighbors. Choosing an optimal bandwidth value is a critical choice because 1) it tells the kernel how big is the neighborhood around the target datapoint in the training set, and 2) it avoids under and over-smoothing [26][15][16]. The smaller bandwidth value is, the closer the regression function will conform to the training data, while using too small bandwidth value is undesirable [23], [26]. In contrast, large bandwidth value produces the smoothest functions that wiggle the least in response to fluctuations in the data. Typical values for bandwidth would lie in the range from 0.2 to 0.5 [27]. To avoid both under and over-smoothing conditions, we used the same set of bandwidth values $b \in \{0.2, 0.3, 0.4, 0.5\}$.[23], [27]

$$K(x_i, x_0) = D\left(\frac{\|x_i - x_0\|}{b}\right) \qquad (3)$$

Where $x_i, x_0 \in \mathbb{R}^p$ and $\|.\|$ is the second norm (Euclidean distance). $D(.)$ is typically a positive real valued function, whose value is decreasing for the increasing distance between the $x_i$ and $x_0$. Usually, $x_0$ is the query example.

Applying the kernel functions with LWR in effort estimation is yet unexplored area. We believe that investigating this area of research might be a worthy topic of research and will be a rich source of future insights into effort estimation. The complete list of all kernel functions used is described in Table 1. All kernel functions produce non-uniform weights except Rectangular function that produces uniform weights. These kernel functions have been used because they are successfully applied in different research domain with promising results.

Table 1. List of all employed kernel functions that are used in this research, Consider $h = \frac{\|x_i - x_0\|}{b}$

| Kernel type | Equation | Kernel type | Equation |
|---|---|---|---|
| Rectangular (Uniform) | $K(h) = 0.5$ | Triweight | $K(h) = \frac{35}{32}(1-h^2)^3$ |
| Epanechnikov | $K(h) = \frac{3}{4}(1-h^2)$ | Biweight | $K(h) = \frac{15}{16}(1-h^2)^2$ |
| Tricube | $K(h) = \frac{70}{81}(1-h^3)^3$ | Cosine | $K(h) = \frac{\pi}{4} cos\left(\frac{\pi}{2} \cdot h\right)$ |
| Gaussian | $K(h) = \frac{1}{\sqrt{2\pi}} e^{\frac{-1}{2}h^2}$ | Logistic | $K(h) = \frac{1}{e^h + 2 + e^{-h}}$ |
| Triangle | $K(h) = 1 - |h|$ | Sigmoid | $K(h) = \frac{2}{\pi} \cdot \frac{1}{e^h + e^{-h}}$ |

## 3. Related Work

Software effort estimation has been intensively studied in the past four decades [3], [4], [29]–[33], where hundreds of research papers have been published to solve various challenges that are mainly related to enhancing the accuracy of estimation. Software effort estimation can be classified into two main groups [31]: model-based methods and expert judgment. Methods that are based on expert judgement is a method that relies on the estimating from similar experienced cases [7]. There are two ways to apply expert judgment: implicitly through informal meeting of experts or explicitly through following a specific approach such as Delphi [7]. On the other hand, model-based methods do not rely solely on expert judgment, but they can adopt algorithms to map the relationship between input variables and effort [34][35]. The latter approach is classified into 1) Parametric approach [36], or 2) Induced approach such as using Machine learning algorithms [7], [8], [43]–[45], [29], [31], [37]–[42].

This list is hardly complete. Elsewhere [1], [46], [47], we have examined all the different kinds of studies that build estimation models based on local data methods in the literature. There is an example of successful methods that integrates both regression with *k*NN to adjunct each other. This can be applied in situations when local learning is required to increase its accuracy. In the literature, little work has been made on examining locality for software estimation. Locality-based approaches are methods that use most similar training projects to produce estimation for new project under test. As is known, the datasets used in effort estimation are heterogeneous and small in size. Thus, locality-based methods on these datasets perform better than other methods that do not adopt locality [1], [12], [13], [47]. Examples of Locality-based methods are feature selection, classification, and clustering algorithms. Different studies adopted locality in their prediction models has reported improvements in accuracy. In other work of Gallego et al. [13], EM clustering algorithm was used to divide training projects into homogeneous projects, and then the regression models were constructed for each cluster. They achieved improvements in terms of MMRE[1]. Another approach by Kocaguneli et al. [12] used multiples decision tree models obtained from training projects. They consider locality by using modified regression trees and they reported improvements over traditional estimation by analogy. Regression trees can be seen as rules to separate examples to group of similar examples based on their feature values. Regression trees are created by considering not only the existing input features of the training examples, but also the impact of the input feature values on the dependent variable. This is a potential advantage over clustering approaches, which typically separate data according to their input features only. Minku et al. [1] investigated the effect of adopting locality on the construction of ensembles learning methods. They concluded that the accuracy of ensemble effort estimation could be improved by adopting locality methods such as clustering algorithms. Azzeh et al. [9] examined the performance of class decomposition to build local data that help in better classification for the purpose of effort estimation based on use case points. However, literature reviews lack comprehensive studies comparing different locality approaches. So, it is not known whether regression trees would be more adequate locality approaches than cluster-based approaches or which of them should be further exploited with the aim of improving effort estimation. The summary of most related works is presented in Table 2.

---
[1] MMRE stands for Mean Magnitude of Relative Errors.

Moreover, the work of [1], [48] have reported improvements in prediction performance when local methods are used to predict effort estimation based on completed projects similar to those to be estimated. Other studies reported accuracy improvements when using "within company" data with their effort estimation models. But they commented that sometimes the "within company" training set examples are believed to be not large enough to efficiently represent the whole population of projects. Thus several effort estimation studies explored the use of "cross-company" models that eventually resulted an inefficient effort estimation models [16]. For instance, several studies expanded "within company" training sets by using training examples from "cross-company" [49] which are then used to build machine learning models for predictions in the context of "within-company". Models trained on "within company" training examples have been reported to achieve better or even similar predictive performance than those previously mentioned in this section [1]. Another recent study investigated relevancy filtering and showed that excluding "cross-company" projects dissimilar from the projects used in prediction often resulted in Cross Company models that can achieve similar predictive performance to "within-company" models.

Table 2. Summary of related work

| ref | Objective | findings |
|---|---|---|
| [1] | Evaluating and analyzing the adequacy of different locality approaches with ensemble learning with purpose to improve effort estimation accuracy. | Combining locality with ensemble lead to competitive results in software effort estimation. |
| [9] | This paper proposed a new method to predict productivity from UCP environmental factors by applying classification with decomposition technique. | Using class decomposition with productivity classification can lead to adequate results. |
| [10] | They used Tabu search (TS) to configure Support Vector Machine (SVR) for single and cross web project effort estimation | The use of TS allowed us to automatically obtain suitable parameters' choices required to run SVR. Moreover, the combination of TS and SVR significantly outperformed all the other techniques. |
| [11] | They investigate Dycom approach using six clustering methods and three automated tuning procedures is performed, to check whether clustering with automated tuning can create well performing Cross Company splits. | the proposed online supervised tuning procedure was generally successful in enabling a very simple threshold-based clustering approach to obtain the most competitive Dycom results |
| [12] | to give insights on the use of multiple learners over single and multiple datasets. | multiple learners induced on single dataset do not produce significantly better results. |
| [15] | This paper examines the impact of data locality approaches on productivity and effort prediction from multiple UCP variables. | The results confirmed that the prediction models that are created based on local data surpass models that use entire data. |
| [43] | This paper aims to comparatively evaluate local versus global lessons learned for effort estimation and defect prediction | They concluded that when researchers attempt to draw lessons from some historical data source, they should ignore any existing local divisions into multiple sources, and cluster across all available data, then restrict the learning of lessons to the clusters from other sources that are nearest to the test data. |
| [44] | They applied Expectation maximization clustering technique to alleviate the problem of heterogeneity in software datasets | Results demonstrate the adequacy of using clustering techniques for software effort datasets. |

4. Data

The accuracy of the different prediction models was evaluated using seven publicly available datasets from PROMISE [50] data repository which consists of datasets donated by various researchers around the world. The choice of public datasets facilitates replicating and feasibility of our study[51]. These datasets are heavily used in software effort estimation research, and they exhibit different characteristics as shown in Table 2. The datasets come from PROMISE are: Desharnais, Kemerer, Albrecht, Maxwell, China, Telecom and Nasa datasets. The employed datasets typically contain a unique set of features that can be categorized according to four classes [29]: size features, development features, environment features and project data. Albrecht contains 24 projects where 18 projects were written in COBOL and 4 projects were written in PL1 and the rest were written in other languages. Two projects have an effort value of more than 100,000 hours which is twice as large as the third largest project. These extreme projects may have a negative impact on the prediction, but it was preferable to keep them. The Kemerer dataset contains 15 software projects described with 7 features. Independent features are represented by 2 categorical and 4 numerical features. The Desharnais dataset originally consisted of 81 software projects collected from Canadian software houses. This dataset is described by 12 features, two dependent features which are the duration and effort measured in '*person-hours*', and 10 independent features. Unfortunately, 4 projects out of the 81 had missing values so it was omitted as it is a misleading estimation process as discussed later. This data pre-processing resulted in 77 complete software projects. The Maxwell dataset contains 62 projects described in 27 features and collected from one of the largest commercial banks in Finland between 1985 and 1993. The China dataset contains 499 software projects described with 16 features and two dependent features Effort and Duration. All features in the China dataset are numeric. The Telecom and Nasa datasets are relatively small dataset with 18 and 22 projects respectively with 2 features. Telecom was collected from one division of a large telecom company.

Table 2 shows the descriptive statistics of such datasets. From these statistics we can conclude that datasets in the area of software effort estimation share relatively common characteristics. They often have a limited number of observations that are affected by multi-collinearity and outliers. But they have a different unit of measure for effort and size. Usually, software effort is measured in months, but for some reasons, effort can be measured in hours to be more accurate. It can be seen that the median of effort for all datasets are much smaller than the mean indicating that is skewed to the right rather than normal. Notably, all datasets have positive skewness efforts ranging from 0.57 to 4.4 which indicates that the effort of each dataset is not normally distributed and is a challenge to develop an accurate estimation model. Finally, the range of effort variable for all datasets indicating a large variability of the projects collected.

Table 3 Statistical properties of the employed dataset

| Dataset | Feature | #instances | Software Size Unit | Effort Data | | | | | |
|---|---|---|---|---|---|---|---|---|---|
| | | | | Unit | min | max | mean | median | skew |
| Albrecht | 7 | 24 | Function Points | Hours | 1000 | 105,000 | 22,000 | 12,000 | 2.2 |
| Kemerer | 7 | 15 | LOC | Months | 23.2 | 1107.3 | 219.2 | 130.3 | 2.76 |
| Nasa | 3 | 18 | LOC | Months | 5 | 138.3 | 49.47 | 26.5 | 0.57 |
| Desharnais | 12 | 77 | Function Points | Hours | 546 | 23940 | 5046 | 3647 | 2.0 |
| China | 18 | 499 | Function Points | Hours | 26 | 54620 | 3921 | 1829 | 3.92 |
| Maxwell | 27 | 62 | Function Points | Hours | 583 | 63694 | 8223.2 | 5189.5 | 3.26 |
| Telecom | 3 | 18 | Files | Months | 23.54 | 1115.5 | 284.33 | 222.53 | 1.78 |

All datasets have been initially pre-processed to avoid the negative effect of missing values. Using these missing values could lead to high degree of inaccuracy in effort estimation. There are various techniques for handling missing values such as 1) deletion, 2) imputation and 3) prediction. The first technique implies deleting the row or column that contains missing values. The second technique implies replacing the missing values with the mean or median of the column values. The third technique implies building a prediction model to predict missing values. Each of these approaches has its pros and cons, for example the deletion technique may reduce the number of instances to a level where a robust model cannot be built. The second technique can cause data leakage and does not affect the covariance the covariance between features. Finally, the third technique can increase the complexity of the model and may fall within the bias of a particular feature. However, since the number of missing values is considered to be very small for each data set, we use the row deletion technique [52]. For example, 4 projects in Desharnais dataset include missing values in their features, therefore they have been excluded. Regarding categorical features, they have been converted into numeric scale using one hot encoder algorithm which transforms each categorical feature with $m$ categories possible values into $m$ binary features, one of them is 1, and all others are 0. Moreover, it is necessary to remove any after-the-event features from the data (other than the effort). For example, the feature duration in Desharnais dataset is removed as it is considered dependent feature. Finally, since each feature has a difference scale (i.e., the range is different), all numeric input features in each data set were re-scaled using a min-max approach such that all numeric feature values are in the range [0, 1] as shown in equation 4.

$$\bar{x} = \frac{x - \min(X)}{\max(x) - \min(x)} \quad (4)$$

Where $x$ is the value before scaling, $\bar{x}$ is the value after scaling and $X$ is the list of all feature values.

In order to better analyze the relationship between the performance of LWR+kernel functions and dataset characteristics we categorized datasets into three classes based on their dimension (i.e. number of rows and columns):
1. *Low*: Albrecht, Kemerer, Nasa and Telecom datasets.
2. *Medium*: Desharnais, Maxwell datasets.
3. *Large*: China dataset.

5. Evaluation Measures

Evaluation measures typically comment on the success of a prediction model. In the field of effort estimation, there are different evaluation measures have been used such as Magnitude Relative Error (MRE) and its aggregated forms: Mean Magnitude Relative Error (MMRE) and Median Magnitude Relative Error (MdMRE). Multiple studies raised concerns about bias produced by MRE measure as it is unbalanced and generates asymmetry distribution [53], [54]. Thus, it is not recommended using MRE and their derived measures to compare between prediction models or to evaluate a single model. Other measures such as mean of Balanced Relative error (MBRE), mean of Inverted Balanced Relative Error (MIBRE) are less sensitive to bias but are not widely used. However, using multiple evaluation measures could also confuse the interpretation of the results and produce contradictions in the findings. Therefore, we favor using one measure that is frequently used to validate effort estimation models and distinguished as unbiased measure, which is Mean Absolute Errors (MAE) [4], [25]. The MAE computes the average of absolute errors between the actual effort of a particular project ($e_i$) and the predicted effort of that project ($\bar{e}_i$) as shown in equation 5. This measure should be as small as possible because large deviation will have opposite effect on the development progress of the new software project.

$$MAE = \frac{\sum_{i=1}^{n}|e_i - \bar{e}_i|}{n} \quad (5)$$

Where $n$ is the number of projects in the dataset.

The second evaluation measure is Scott-Knott analysis, which is a statistical procedure of multiple comparison based on the idea of clustering where the criterion of clustering is a test of significance based on their ranks [55]. The Scott-Knott procedure essentially follows and uses one-way analysis of variance (one-way ANOVA) of MAE distribution, which tests the null hypothesis that the kernel functions being compared are statistically indifferent against the alternative hypothesis that states that the kernel functions can be partitioned into subgroups. The reason behind the use of Scott-Knott technique is its ability to separate the kernel functions into non-overlapping groups based on the significance test. Furthermore, we used one-way ANOVA to determine if there were any significant differences between the means of three or more independent (unrelated) groups. This test will be used to find statistical differences between all kernel functions with respect to different criteria.

6. Methodology

The main objective of this paper is to investigate the main configuration parameters (i.e., kernel function, bandwidth and polynomial degree) of LWR method for effort estimation. Figure 2 shows our experiment procedure. Mainly, we use 7 datasets, 10 kernel functions, 4 bandwidth values and 3 polynomial degree values which resulted in 7 × 10 × 4 × 3 = 840 different LWR configuration. In other words, for each LWR model we must choose one kernel function, one bandwidth, and one degree value from:

- Kernel type $f \in \begin{Bmatrix} Biweight, Cosine, Epanechnikov, Gaussian, Logistic, \\ Rectangular, Sigmoid, Triangular, Tricube, Triweight \end{Bmatrix}$
- Bandwidth $b \in \{0.2, 0.3, 0.4, 0.5\}$
- Polynomial degree $d \in \{1, 2, 3\}$

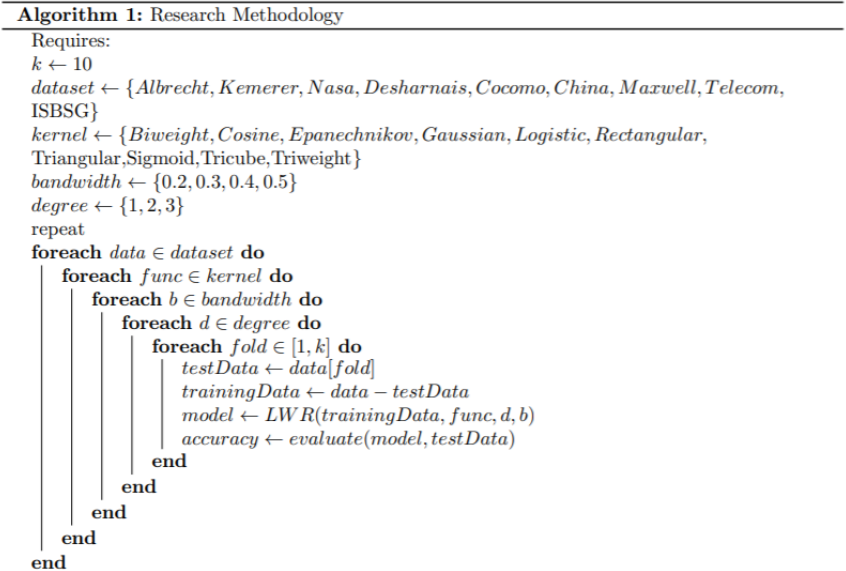

Figure 2. Experimental Process

For each experiment, a combination of different configurations is used to construct LWR model and apply it to each of the seven datasets. For each experiment, we use repeated l0-Folds cross validation which is repeated 10 times. Using this approach attempts to reduce bias when choosing training data within 10-folds cross validation. Basically, the entire dataset is divided into 10 subsets. In each run, one subset is used for testing and the remaining subsets are used for training. This procedure is repeated ten times until all subsets are used for testing. Accuracy measures are recorded for each experiment. It is important to make sure that the LWR variants are constructed based on the same training data to avoid bias in predictions as shown in Figure 2.

The obtained results are presented in different formats including ranking and visualization. Firstly, we discuss the results in terms of MAE aggregation. We focus on MAE because it is a common and unbiased evaluation measure, in addition it is easier for interpretation. Secondly, we rank, and cluster kernel functions based on MAE distribution using Scott-Knott analysis. Scott-Knott cluster analysis is used for statistical comparison of all kernel functions based on MAE distribution and then cluster them into homogenous subgroups where each subgroup contains significantly indifferent kernel functions. For sake of our analysis, we rank and cluster the LWR variants based on different criteria. In our first criterion of data analysis, we aggregate overall MAE across all experiment conditions (i.e., datasets, bandwidth, and degree values). In the second and third criteria, we examined the impact of degree and bandwidth parameters on the accuracy of kernel functions by aggregating MAE in terms of degree and bandwidth across all datasets. Then Scott-

Knott analysis was performed to determine which specific kernel function significantly performs better than others [56].

## 7. Results and Discussions

As mentioned in the methodology section, a large-scale experiment has been conducted using multiple datasets and kernel functions. Specifically, we used seven datasets and ten kernel functions where for each kernel function we applied different bandwidth and degree parameters. These varieties resulted in 840 experiments. Each constructed LWR model has been evaluated using MAE. For simplicity, throughout the remaining section we use kernel function name instead of mentioning LWR+kernel function name. There are 12 possible variants for each LWR+kernel as result of the intersection between bandwidth and degree parameters.

### 7.1 Overall Aggregation

The first part of our analysis is to show the aggregate results of MAE for all LWR models across all datasets. The MAE aggregation process is illustrated in equation 5. The average of MAE for all kernel functions are presented for each dataset individually in Table 4. In other words, each cell represents the average of all MAE values for all combinations of LWR and kernel functions as explained in equation 5. For example, the intersection between Albrecht and Biweight represents the average of all MAE values for all LWR+Biweight variants that are constructed over Albrecht dataset. The MAE values with boldface denote the most superior kernel function for each dataset. From the table we can observe that four datasets favor Triweight kernel functions, and three datasets favors Biweight. It is clear that both pre-mentioned kernel functions are the most fit with LWR models for effort estimation. Three findings are worth commenting, 1) the results showed that using uniform kernel function (i.e., Rectangular) in general offers no improvement over non-uniform kernel function which does outperform uniform kernel with an exception to three non-uniform kernel functions (Gaussian, Sigmoid and Logistics where uniform kernel function outperforms them. 2) The results also reported that Triweight kernel performs well on small datasets, and Biweight can work well over medium and large datasets.

$$MAE_{i,j} = \frac{1}{|b|}\sum_{t=1}^{|b|}\frac{1}{|d|}\sum_{k=1}^{|d|}MAE_{t,k} \quad \forall i \in f, \forall j \in dataset \quad (5)$$

where:
- $i$ represents index of kernel function
- $j$ represents index of dataset
- $t$ represents index of bandwidth from $b$ vector
- $k$ represents index of degree value form $d$ vector

Table 4. Average of MAE for each Kernel, across all degrees and bandwidths

| Kernel | Albrecht | Desharnais | Kemerer | China | Maxwell | NASA | Telecom |
|---|---|---|---|---|---|---|---|
| Biweight | 11.7 | **117.8** | 21.1 | **613.0** | **116.9** | 9.8 | 6.2 |
| Cosine | 13.6 | 144.5 | 19.6 | 663.7 | 144.2 | 10.8 | 6.4 |
| Epanechnikov | 12.6 | 147.6 | 17.8 | 669.7 | 147.5 | 11.0 | 6.5 |
| Gaussian | 21.7 | 397.4 | 17.1 | 969.5 | 369.4 | 19.0 | 31.8 |
| Logistic | 34.8 | 657.4 | 19.6 | 1156.7 | 597.8 | 26.9 | 78.2 |
| Rectangular | 17.8 | 240.6 | 21.3 | 852.3 | 248.8 | 13.3 | 8.1 |
| Sigmoid | 28.4 | 572.9 | 16.6 | 1085.4 | 484.4 | 24.7 | 57.0 |
| Triangular | 11.9 | 127.9 | 18.7 | 631.5 | 128.8 | 9.6 | 6.1 |
| Tricube | 11.8 | 137.8 | 17.4 | 651.7 | 135.8 | 10.0 | 6.3 |
| Triweight | **11.1** | 131.4 | **16.3** | 639.3 | 130.4 | **9.0** | **6.0** |

Next, we examine the MAE distribution for all LWR models over each dataset using interval plot with 95% significant confidence interval of MAE mean as shown in Figure 3.

The interval plot with 95% confidence is used for two reasons: 1) it is a graphical data interpretation technique which allows us to visually assess the difference between multiple groups and build provisional decision. 2) Since we are interested to investigate the significant differences between means of different MAE populations the point mean estimate alone as in Table 4 is not enough to judge that multiple MAE distribution are significantly different. It means that for each such sample, the mean, standard deviation, and sample size are used to construct a confidence interval representing a specified degree of confidence, say 95%. Under these conditions, it is expected that 95% of these sample-specific confidence intervals will include the population mean. We draw interval plot for each dataset because each one has different MAE scale.

From Figure 3 we can observe the following findings:
1. The four Kernel functions (Sigmoid, Gaussian, Logistic and Rectangular) degraded the performance of LWR over almost all datasets. We could not find that any of these algorithms contributes enough to improving accuracy of LWR method. Therefore, the first impression encourages us to ignore these kernels from being used with LWR for effort estimation datasets. Also, we can notice that there is remarkable significant difference between their MAE distribution and other kernel distribution.
   Our samples provide evidence that there is a difference in the average torque resistance values between one or more of the five lots.
2. A one-way ANOVA test among all 10 kernel functions, on each dataset separately, revealed rejection of the null hypothesis. This provides evidence that there is a difference in the average MAE values between one or more kernel functions. However, we have one exception on the Kemerer dataset where kernel functionality is not significantly different. There is no clear reason why this would only happen on this dataset. One reason could be that there are two outliers (i.e. extreme projects) in these datasets which may produce local regression models biased to those projects.
3. Biweight and Triweight kernel function can work best over seven datasets (except Kemerer dataset) as they have minimum MAE distribution. However, we could not find significant differences between their MAE distribution and other four kernel functions (Epanechnikov, Tricube, Cosine and Triangular)

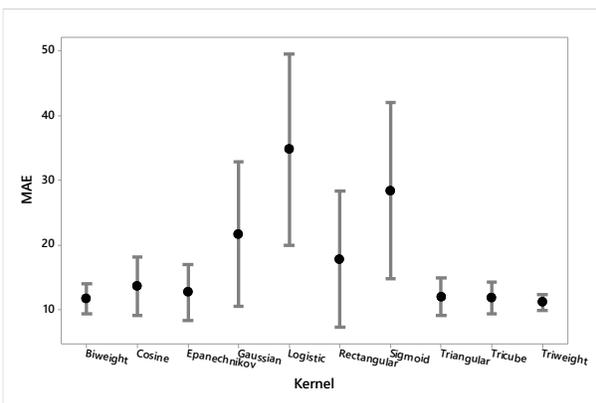
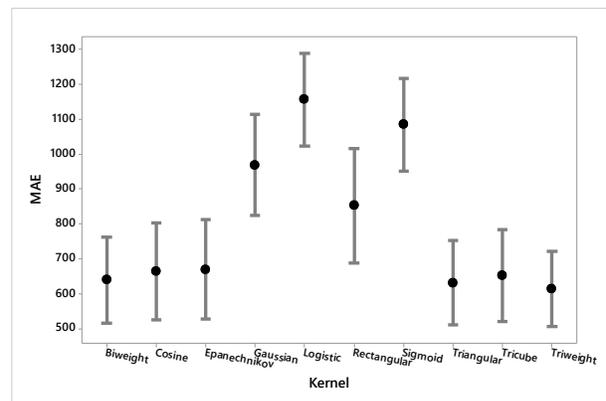

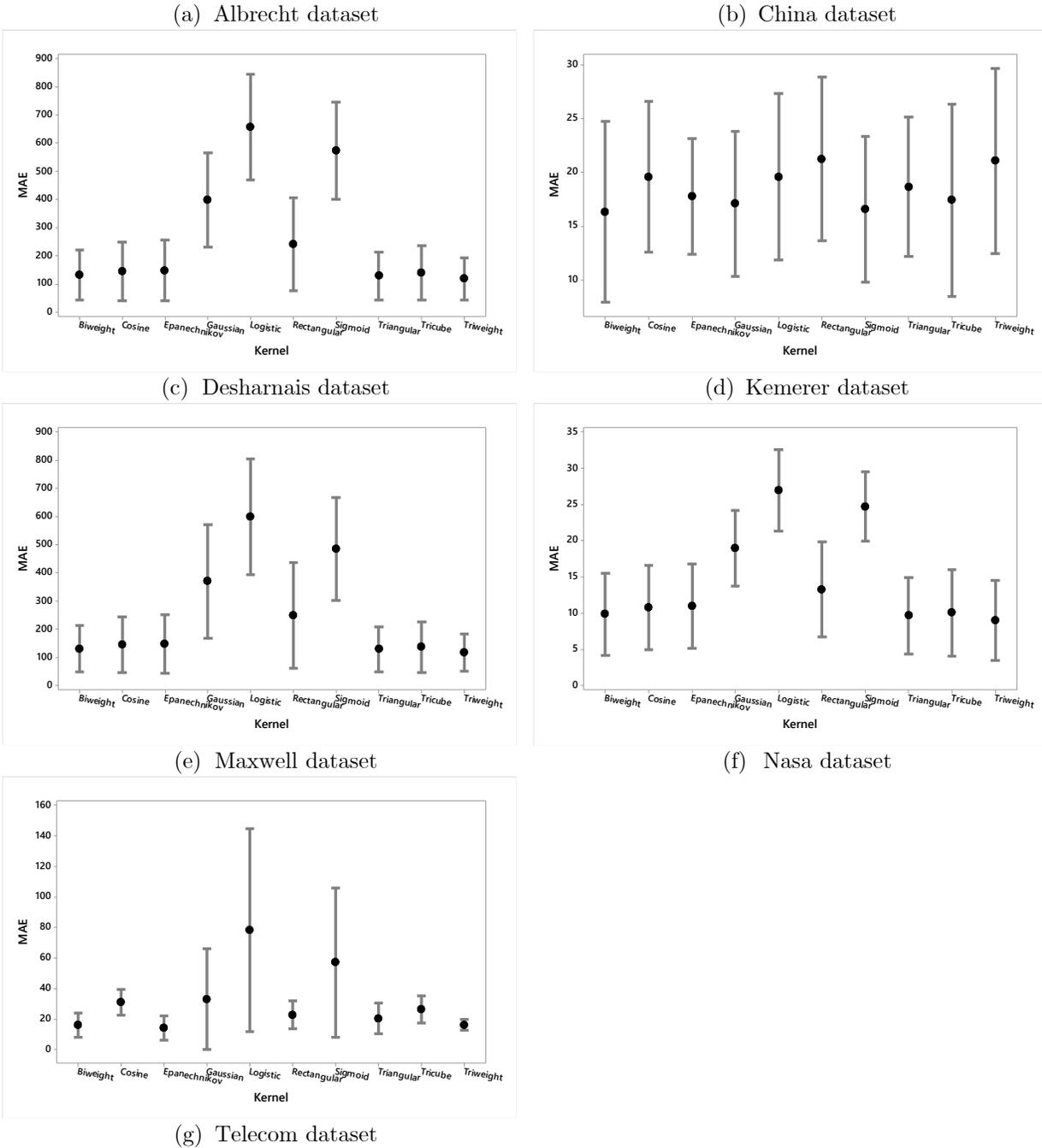

Figure 3. Interval plots of 95% confidence interval of MAE distribution for all LWR variants over each individual dataset.

In addition to the above analysis, we consult Scott-Knott analysis to significantly cluster and rank all LWR variants based on their MAE distribution. Since every dataset presents different MAE range, we scale the MAE of all LWR variants over all each dataset individually using Box cox function. The kernel functions on the right-hand side are the higher-ranked functions, while the functions on the left are the last-ranked functions as shown in Figure 4. Functions are grouped based on the significance test into multiple groups where the functions within the same group are significantly indifferent. However, all kernel functions are clustered into three significant groups.

From this figure we can observe that Triweight and Biweight (surrounded by grey border) are the most significant ranked functions across all experiment conditions. Moreover, we can see that Logistic, Gaussian and Sigmoid are the last ranking across multiple experiments. Notably, Epanechnikov, a common kernel function, is ranked in the middle with insignificant difference from the Triangle, Biweight and Tricube kernel functions. Regarding distribution of rankings, we can notice that all kernel functions show relatively large deviation in their ranks, which demonstrate that there is instability in ranking of kernel functions within a particular accuracy measure. Surprisingly, the rank distribution for the last ranked kernel functions presents less deviation in comparison with top ranked kernel functions which tells us that the last ranked functions are also stable but not accurate. Based on these findings we can rank the kernel functions in the following definite order: Triweight, Biweight ≻ Tricube, Triangle, Cosine, Epanechnikov, Rectangular ≻ Gaussian, Sigmoid, Logistic. The ≻ symbol means the precedent kernel is preference over the consequent kernel. Therefore, the conclusion drawn from this part recommends using Triweight and Biweight with LWR which could improve accuracy of local regression for effort datasets.

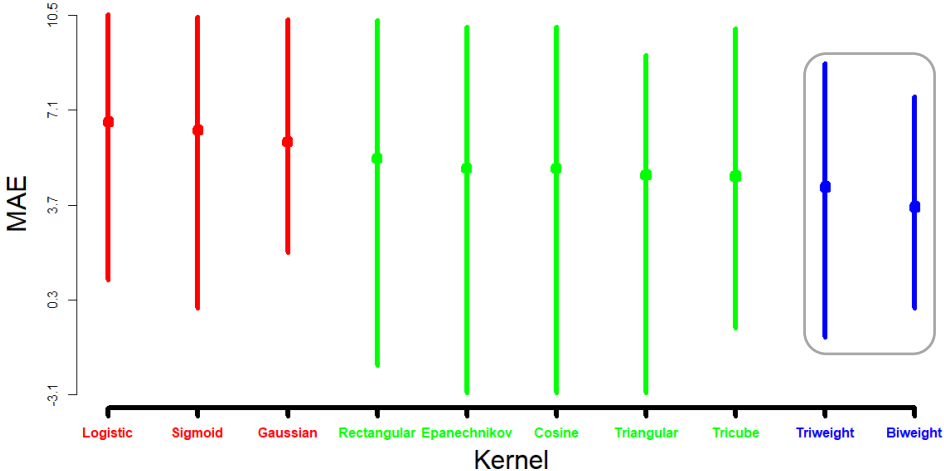

Figure 4. Scott-Knott analysis of ranks for all kernel functions over all experiment conditions

## 7.2  Examining the impact of Polynomial degree

This section examines the impact of polynomial degree parameter on the accuracy of LWR. The process of analyzing MAE using interval plots is described in Figure 4. The degree of polynomial that is used to create LWR model can take one of three values (1, 2 or 3). Figure 6 shows the relationship between average of MAE and degree of LWR polynomial for each kernel function and dataset. In other words, each scatter plot represents the intersection between one LWR variant as column and one dataset as row. The complete source data for this analysis can be found in Table A1 in the appendix section. The main finding is that all datasets favor degree=3 except Albrecht and Kemerer datasets where they reported conflicting results. The Albrecht dataset would seem favor degree=1 whereas Kemerer dataset has no stable favorite degree. On the other hand, if we look at the favorite of each kernel across all effort datasets, we can notice that all of them favor degree=3. Hence, there is strong indication that constructing LWR polynomial with degree= 3 would work well with almost all effort datasets. In summary, we can confirm that using degree =3 can produce good and accurate results for medium and large-scale datasets.

```
Algorithm 2: Interval plot Analysis with respect to degree
AllMAE=readAllMAEs()
dataset ← [Albrecht, Kemerer, Nasa, Desharnais, Cocomo, China, Maxwell, Telecom]
kernel ← [Biweight, Cosine, Epanechnikov, Gaussian, Logistic, Rectangular,
    Triangular,Sigmoid,Tricube,Triweight]
bandwidth ← [1, 2, 3, 4]
degree ← [1, 2, 3]
MAE= ← [ ]
foreach k ∈ Kernel do
    foreach s ∈ Dataset do
        foreach d ∈ Degree do
        | MAE[k] ← MAE.append(AllMAE(s, d))
        end
    end
end
DrawIntervalPlot(MAE)
```

Figure 5. Procedure of constructing Interval plots with respect to degree value

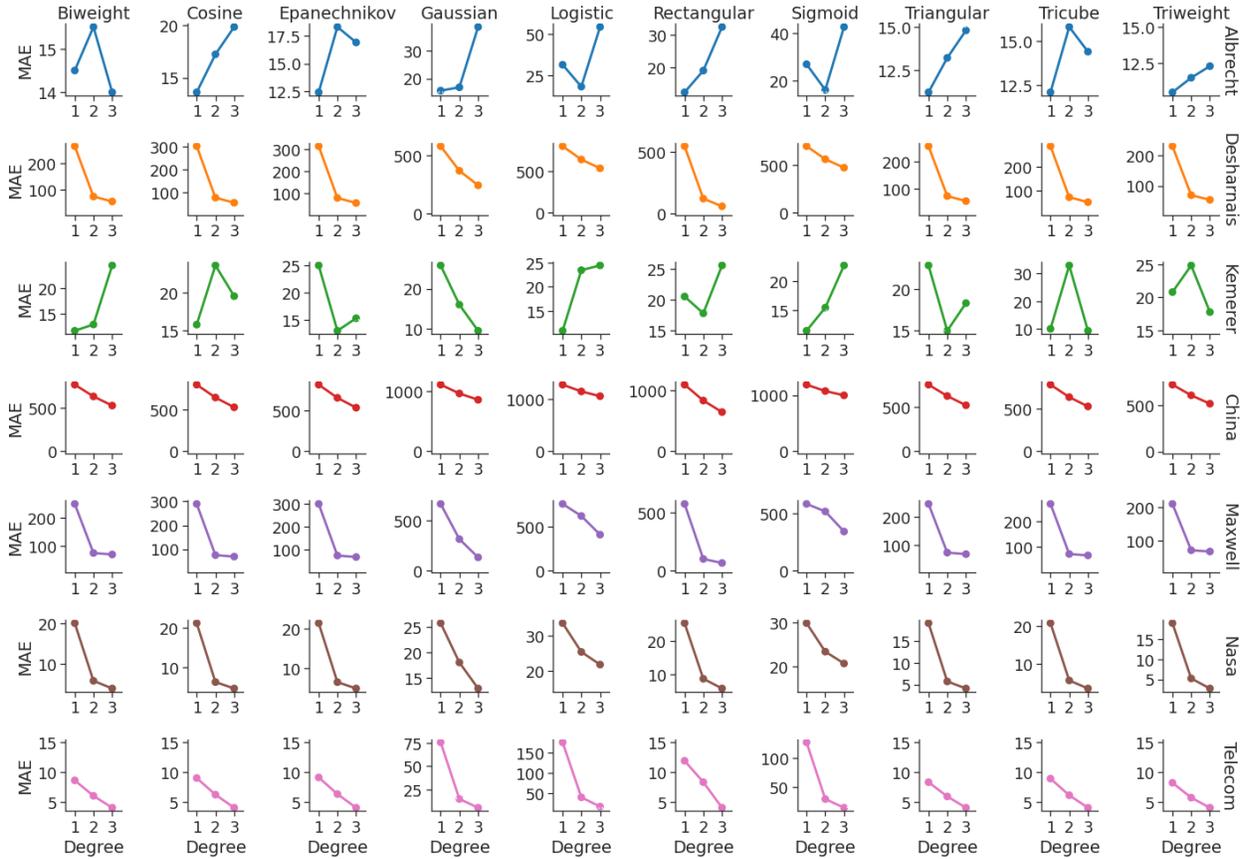

Figure 6. Relationship between average of MAE and degree of LWR polynomial across each dataset

These results are also confirmed by the interval plots of 95% confidence interval of MAE distribution as shown in figure 7. In this Figure, we summarized the MAE distribution over each kernel function with respect to each degree value. For each LWR+kernel combination we draw three interval plots corresponding to each degree value used in that combination. Surprisingly, it is clear that using LWR with degree 3 produce smaller MAE distribution with minimum mean of MAE over all datasets.

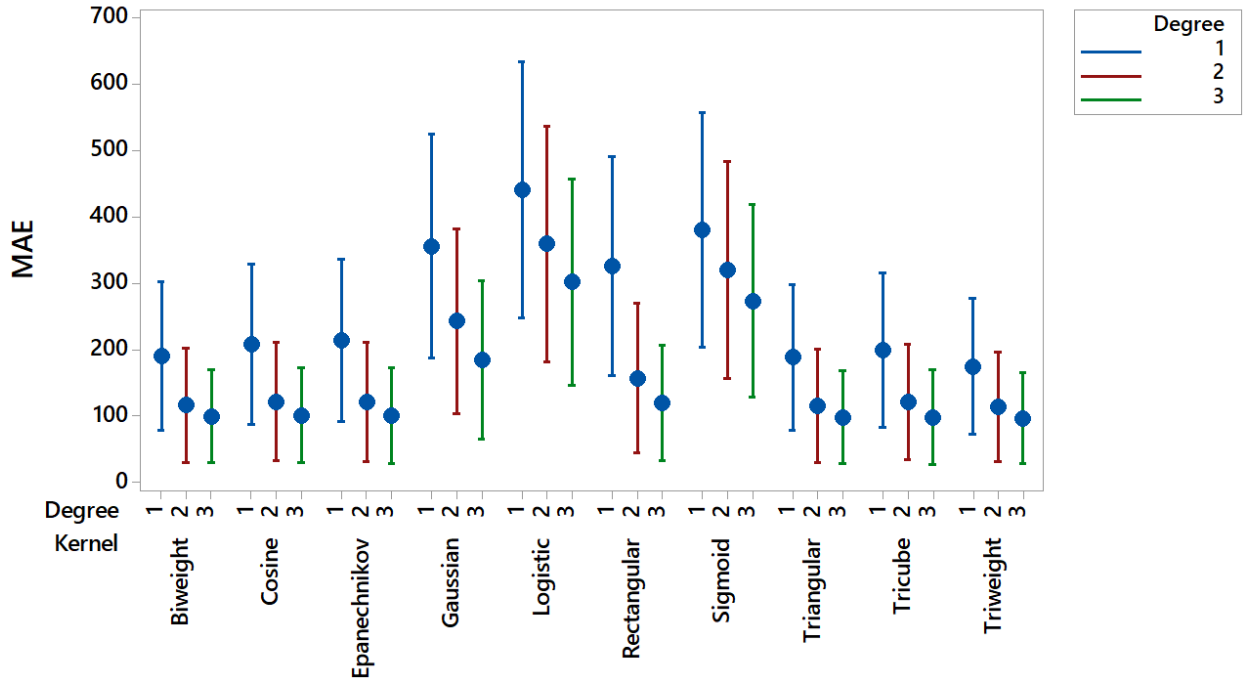

Figure 7. Interval plot of 95% confidence of MAE Distribution for all LWR variants.

One-way ANOVA was performed to compare the MAE value of three polynomial degree variants. The test revealed that the null hypothesis is rejected with respect to each kernel function. This provides evidence that not all MAE means are the same for every variant. Therefore, we can conclude that using degree=3 is significantly different and superior.

### 7.3 Examining the impact of kernel bandwidth

This part of analysis examines the impact of kernel bandwidth on predictive accuracy of LWR based effort estimation. The process of analyzing MAE using interval plots with respect to bandwidth is described in Figure 8. The bandwidth of a kernel function specifies the set of nearest neighbors that will be involved in building local regression models. The bandwidth values that we have used are changed from 0.2 to 0.5 as recommended by [27]. Basically, we aggregate MAE distribution of all kernel functions according to bandwidth values. The relationship between average of MAE and bandwidth of LWR kernel function for each dataset is shown in Figure 9. Each scatter plot represent intersection between one LWR variant as column and one dataset as row. The complete source data for this analysis can be found in Table A2 in the appendix section. The main finding is that all datasets favor combination kernel with bandwidth =0.2 except Albrecht and Kemerer datasets. The Albrecht dataset would seem have no clear favorite bandwidth, but in general bandwidth =0.2 and 0.5 can work well with clear improvements, whereas Kemerer dataset has no stable favorite bandwidth. On the other hand, if we look at the favorite of each kernel across all effort datasets, we can notice that all of them favor bandwidth=0.2. Hence, there is strong indication that constructing LWR polynomial with bandwidth=0.3 irrespective degree value would work well with almost all effort datasets. In summary we can confirm that using bandwidth =0.2 can produce good and accurate results for medium and large-scale datasets in addition to some small datasets with small dimension such as Nasa and Telecom.

**Algorithm 3:** Interval plot Analysis with respect to Bandwidth

```
AllMAE=readAllMAEs()
ddataset ← [Albrecht, Kemerer, Nasa, Desharnais, Cocomo, China, Maxwell, Telecom]
kernel ← [Biweight, Cosine, Epanechnikov, Gaussian, Logistic, Rectangular,
  Triangular,Sigmoid,Tricube,Triweight]
bandwidth ← [1, 2, 3, 4]
degree ← [1, 2, 3]
MAE= ← [ ]
foreach k ∈ Kernel do
    foreach s ∈ Dataset do
        foreach b ∈ bandwidth do
         |  MAE[k] ← MAE.append(AllMAE(s, b)
        end
    end
end
DrawIntervalPlot(MAE)
```

Figure 8. Procedure of constructing Interval plots with respect to bandwidth value

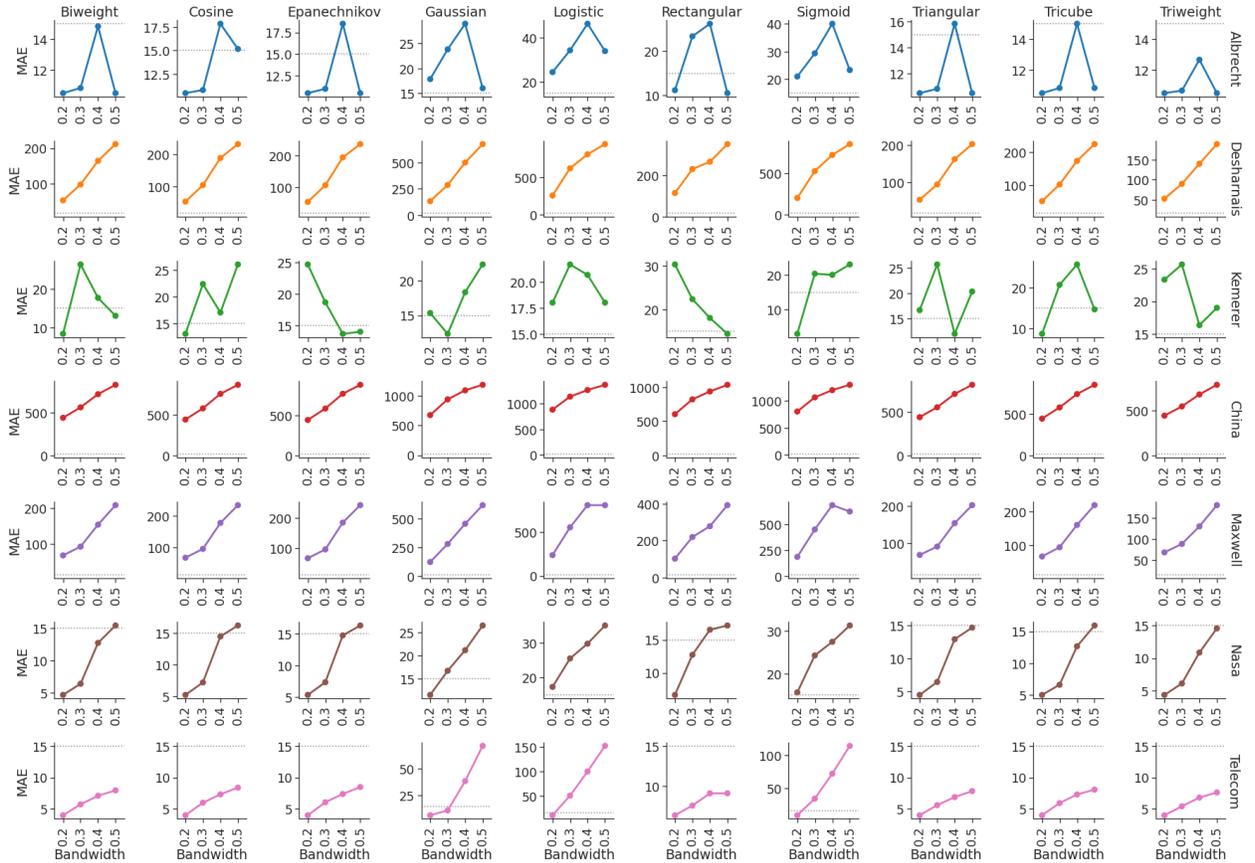

Figure 9. Relationship between average of MAE and bandwidth of kernel across each dataset

These results are also confirmed by the interval plots of 95% confidence interval of MAE distribution as shown in figure 10. In this Figure, we summarized the MAE distribution over each kernel function with respect to each bandwidth value. For each LWR+kernel combination we draw three interval plots corresponding to bandwidth value used in that combination. It is clear that using LWR variants with bandwidth=0.2 produce smaller MAE distribution with minimum mean of MAE over all datasets.

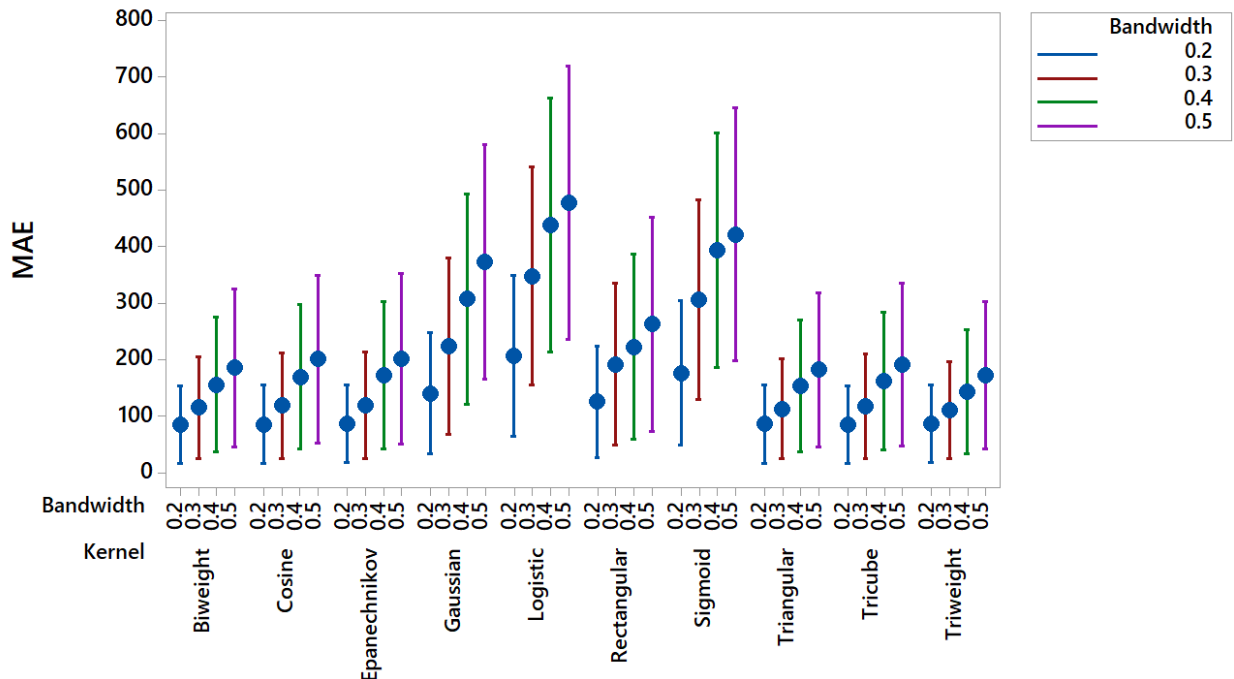

Figure 10. Interval plot of 95% confidence of MAE Distribution for all LWR variants

One-way ANOVA was performed to compare the MAE value of four bandwidth variants. The test revealed that the null hypothesis is rejected with respect to each kernel function. This provides evidence that not all MAE means are the same for every variant. Therefore, we can conclude that the use of bandwidth=0.2 is significantly different and superior.

## 8. THREATS TO VALIDITY

The threats to validity of this study are divided into internal and construct threats. The main internal threat is the choice of bandwidth and degree values. Actually there is no clear recommendation about choosing best range for bandwidth, however we have selected the range from 0.2 to 0.5 to avoid both under and over-smoothing conditions as recommended by [23]. On the other hand, choosing degree values from 1 to 3 allows us to test both linear and nonlinear polynomial as form of regression for LWR models. Increasing degree of polynomial will eventually increase time of learning and thus increasing complexity of the constructed model. We believe choose degree value up to 3 is more reasonable to avoid overfitting and reduce model complexity [24]. Regarding construction threat, we mention the choice of validation approach. In this paper, the repeated 10 folds cross validation which is used to validate all LWR models. Although this approach has not been recommended by [57], but we used this approach for two reasons: first, it is costly less intensive than leave one out cross validation especially for large datasets. As mentioned early in the introduction that LWR is a lazy learning algorithm and takes extraordinarily huge time to construct final model. Second, this approach produces errors very close to leave one cross validation for large datasets. Finally, the datasets employed in this study are not all datasets available, but we favored to use the most common public datasets in order to facilitate replication of our experiments.

9. Conclusions

In this research we examine the use of different kernel functions with LWR for the problem of software effort estimation. We conducted experiments with various kernels as well as bandwidths and degrees. From vast number of experiments there were hardly any case where uniform kernel that produces uniform weighting are accurate than kernels that produce non-uniform weighting for LWR model. In addition, we did not find stability in these kernels across different parameter settings. This discourages us to say that a specific kernel function would generate better weighting for LWR model. However, back to the obtained results, among all kernel functions we can say that non-uniform kernels (e.g., Triweight) are more appropriate than uniform kernel for almost all conducted experiments, hence both Triweight and Triangle kernel functions would be the most appropriate kernels for LWR based effort estimation. The most interesting part from our analysis is that bandwidth and degree values do not affect the choice of the kernel. If we were asked to recommend a degree and bandwidth values for LWR model, we can recommend using degree $=3$ and bandwidth $= 0.2$ as suggested by Scott-Knott analysis. Which means that the kernel functions are more accurate when we increase the number of nearest neighbors in the constructing local regression model. Also, using linear polynomial (i.e., when degree $=1$) is better than using nonlinear regression model as confirmed in section 7.2. However, if similar use of kernels is to be adopted, we do not recommend the use of uniform kernel function as a weighting strategy in LWR based effort estimation. Unlike studies in different domains that used kernel functions with nonlinear regression and reported improved accuracy values [18, 47], we do not observe such an effect on software effort datasets. The reason for different results may lie in different dataset characteristics. For instance, the datasets used in other domains are much more densely populated than software effort datasets. Below we revisit and address the proposed research questions:

*RQ1. Is there evidence that non-uniform kernel functions can outperform uniform kernel functions for LWR?*

The results of our experiments do not show any evidence that uniform kernel (i.e., Rectangular) would enhance the accuracy of LWR as confirmed by Friedman's test and Scott-Knott analysis. On the contrary, all other kernel functions efficiently work with LWR and yield much better results than uniform kernel. The overall results did not show stability in ranking non uniform kernel function across multiple experiments. But based on the majority of rankings we have found the Triweight kernel function would give much better accuracy and stability than other non-uniform kernel functions.

*RQ2. What are the interactions between: (i) data set dimensionality (ii) type of kernel (iii) polynomial degree and (iv) kernel bandwidth?*

Changing kernel bandwidths from 0.2 to 0.5 shows a random and insignificant effect on ranking kernel functions. However, the large bandwidth values would give less deviation in ranking than smaller bandwidth values. Changing polynomial degree of LWR from 1 to 3 shows insignificant effect on ranking kernel functions. which is very similar to that of kernel bandwidth change effect. Therefore, applying different polynomial degrees with certain kernel bandwidth did not show certain effect on LWR performance. However, the larger degree would give less deviation in ranking with certain bandwidth values than other smaller degree values. Based on Scott-Knott analysis, we found that using LWR polynomial degree $=3$ and kernel bandwidth $= 0.2$ is the best choice for most kernels. Finally, we have found that using Triweight function can work well with small effort datasets and Biweight function works best with large and medium datasets. However, there is no clear pattern for small effort datasets.

Acknowledgments

Authors are grateful to the Applied Science Private University, Amman, Jordan, for the financial support granted to cover the publication fee of this research article.

Mohammad Azzeh is grateful to the Princess Sumaya University for Technology, Amman, Jordan, for the financial support granted to cover part of the publication fee of this research article.

Appendix 1: Detailed results

Table A1. Average MAE for each Kernel for all degrees

| Kernel | Albrecht | | | Desharnais | | | Kemerer | | | China | | | Maxwell | | | NASA | | | Telecom | | |
|---|---|---|---|---|---|---|---|---|---|---|---|---|---|---|---|---|---|---|---|---|---|
| | 1 | 2 | 3 | 1 | 2 | 3 | 1 | 2 | 3 | 1 | 2 | 3 | 1 | 2 | 3 | 1 | 2 | 3 | 1 | 2 | 3 |
| Biweight | 14.5 | 15.5 | 14.0 | 267.7 | 72.9 | 53.8 | 11.8 | 13.0 | 24.3 | 764.4 | 629.3 | 524.3 | 249.8 | 73.3 | 68.1 | 20.0 | 5.7 | 3.8 | 8.6 | 6.0 | 4.0 |
| Cosine | 13.6 | 17.2 | 19.8 | 303.7 | 76.1 | 53.8 | 15.8 | 23.5 | 19.5 | 807.4 | 651.1 | 532.7 | 290.0 | 74.4 | 68.2 | 21.1 | 6.4 | 4.8 | 9.0 | 6.2 | 4.0 |
| Epanechnikov | 12.4 | 18.3 | 16.9 | 312.3 | 76.8 | 53.8 | 25.0 | 13.0 | 15.3 | 818.1 | 656.3 | 534.8 | 299.5 | 74.7 | 68.2 | 21.4 | 6.5 | 4.9 | 9.1 | 6.3 | 4.0 |
| Gaussian | 15.6 | 16.8 | 38.6 | 581.7 | 368.4 | 242.0 | 25.8 | 16.0 | 9.5 | 1102.3 | 955.7 | 850.5 | 662.2 | 313.3 | 132.5 | 25.8 | 18.1 | 13.0 | 75.5 | 14.6 | 5.2 |
| Logistic | 31.6 | 18.4 | 54.4 | 799.4 | 639.2 | 533.7 | 10.8 | 23.5 | 24.5 | 1274.4 | 1143.5 | 1052.3 | 760.9 | 623.0 | 409.5 | 33.6 | 25.4 | 21.8 | 176.1 | 40.6 | 17.8 |
| Rectangular | 12.5 | 19.1 | 32.4 | 544.9 | 119.5 | 57.3 | 20.5 | 17.8 | 25.5 | 1089.9 | 828.4 | 638.5 | 574.9 | 103.1 | 68.4 | 25.3 | 8.7 | 5.8 | 11.9 | 8.3 | 4.0 |
| Sigmoid | 26.8 | 16.0 | 42.5 | 692.4 | 557.2 | 469.3 | 11.5 | 15.5 | 22.8 | 1187.6 | 1072.8 | 995.9 | 587.0 | 520.9 | 345.2 | 29.9 | 23.4 | 20.7 | 127.0 | 29.5 | 14.6 |
| Triangular | 11.2 | 13.2 | 14.8 | 257.7 | 72.4 | 53.7 | 22.8 | 15.0 | 18.3 | 751.6 | 623.7 | 519.1 | 244.9 | 73.5 | 68.2 | 19.0 | 5.7 | 4.1 | 8.3 | 5.9 | 4.0 |
| Tricube | 12.1 | 15.8 | 14.4 | 285.6 | 74.1 | 53.8 | 10.0 | 33.0 | 9.3 | 786.5 | 639.3 | 529.4 | 265.8 | 73.5 | 68.1 | 20.7 | 5.7 | 3.6 | 8.9 | 6.1 | 4.0 |
| Triweight | 10.5 | 11.5 | 12.3 | 230.7 | 69.0 | 53.6 | 20.8 | 24.8 | 17.8 | 719.8 | 605.0 | 514.0 | 210.8 | 71.8 | 68.1 | 19.0 | 5.2 | 2.7 | 8.2 | 5.7 | 4.0 |

Table A2. Average MAE for each Kernel for all Bandwidth

| Kernel | Albrecht | | | | Desharnais | | | | Kemerer | | | | China | | | | Maxwell | | | | NASA | | | | Telecom | | | |
|---|---|---|---|---|---|---|---|---|---|---|---|---|---|---|---|---|---|---|---|---|---|---|---|---|---|---|---|---|
| | 0.2 | 0.3 | 0.4 | 0.5 | 0.2 | 0.3 | 0.4 | 0.5 | 0.2 | 0.3 | 0.4 | 0.5 | 0.2 | 0.3 | 0.4 | 0.5 | 0.2 | 0.3 | 0.4 | 0.5 | 0.2 | 0.3 | 0.4 | 0.5 | 0.2 | 0.3 | 0.4 | 0.5 |
| Biweight | 10.5 | 10.8 | 14.8 | 10.5 | 52.3 | 97.2 | 164.4 | 211.8 | 8.3 | 26.3 | 17.7 | 13.0 | 442.7 | 564.9 | 719.8 | 829.9 | 68.2 | 92.5 | 153.4 | 207.5 | 4.7 | 6.4 | 12.7 | 15.4 | 4.0 | 5.7 | 7.1 | 7.9 |
| Cosine | 10.5 | 10.8 | 17.8 | 15.2 | 52.4 | 104.0 | 189.1 | 232.6 | 13.0 | 22.3 | 17.0 | 26.0 | 442.7 | 580.9 | 759.7 | 871.5 | 68.2 | 96.1 | 178.2 | 234.3 | 5.2 | 7.2 | 14.4 | 16.2 | 4.0 | 6.0 | 7.3 | 8.4 |
| Epanechnikov | 10.5 | 11.0 | 18.5 | 10.5 | 52.2 | 105.6 | 195.1 | 237.7 | 24.7 | 18.7 | 13.7 | 14.0 | 442.6 | 585.0 | 769.6 | 881.7 | 68.1 | 96.9 | 184.1 | 240.8 | 5.4 | 7.4 | 14.8 | 16.3 | 4.0 | 6.1 | 7.4 | 8.5 |
| Gaussian | 17.8 | 23.8 | 29.0 | 16.0 | 132.0 | 285.8 | 498.4 | 673.2 | 15.3 | 12.3 | 18.3 | 22.3 | 672.3 | 936.4 | 1088.4 | 1180.9 | 124.2 | 280.3 | 455.8 | 617.2 | 11.4 | 16.7 | 21.2 | 26.5 | 6.6 | 11.1 | 38.3 | 71.0 |
| Logistic | 24.5 | 34.3 | 46.3 | 34.0 | 256.2 | 620.6 | 807.8 | 945.0 | 18.0 | 21.7 | 20.7 | 18.0 | 878.7 | 1130.5 | 1258.0 | 1359.6 | 236.0 | 550.7 | 802.5 | 802.0 | 17.3 | 25.5 | 29.8 | 35.0 | 10.3 | 50.5 | 99.7 | 152.2 |
| Rectangular | 11.2 | 23.3 | 26.2 | 10.5 | 114.3 | 230.2 | 266.1 | 351.7 | 30.3 | 22.3 | 18.0 | 14.3 | 605.4 | 824.5 | 940.4 | 1038.7 | 102.0 | 220.0 | 279.1 | 394.1 | 6.6 | 12.7 | 16.5 | 17.2 | 6.4 | 7.6 | 9.1 | 9.1 |
| Sigmoid | 21.0 | 29.3 | 40.0 | 23.3 | 201.1 | 523.7 | 718.3 | 848.7 | 13.0 | 20.3 | 20.0 | 23.0 | 800.0 | 1059.1 | 1192.3 | 1290.2 | 185.1 | 448.5 | 682.5 | 621.5 | 15.6 | 24.3 | 27.5 | 31.3 | 8.5 | 33.8 | 71.7 | 114.1 |
| Triangular | 10.5 | 10.8 | 15.8 | 10.5 | 52.2 | 93.9 | 162.3 | 203.3 | 16.7 | 25.7 | 12.0 | 20.3 | 442.6 | 556.8 | 710.4 | 816.1 | 68.1 | 90.7 | 154.0 | 202.6 | 4.5 | 6.4 | 12.9 | 14.7 | 4.0 | 5.6 | 6.9 | 7.8 |
| Tricube | 10.5 | 10.8 | 15.0 | 10.8 | 52.2 | 102.2 | 173.2 | 223.7 | 8.7 | 20.7 | 25.7 | 14.7 | 442.7 | 577.1 | 737.8 | 849.3 | 68.1 | 95.1 | 160.9 | 219.0 | 5.0 | 6.6 | 12.6 | 15.9 | 4.0 | 5.9 | 7.3 | 8.1 |
| Triweight | 10.5 | 10.7 | 12.7 | 10.5 | 52.2 | 89.4 | 140.1 | 189.4 | 23.3 | 25.7 | 16.3 | 19.0 | 442.6 | 545.6 | 678.4 | 785.2 | 68.1 | 88.2 | 130.4 | 181.0 | 4.3 | 6.1 | 10.8 | 14.6 | 4.0 | 5.4 | 6.8 | 7.6 |